# BI-METRIC EINSTEIN GENERAL RELATIVITY CONTAINS EXPONENTIAL METRICS WHICH DYNAMICALLY PRESERVE BI-METRIC "LIGHT CONE CAUSALITY"(*)

Darryl J. Leiter[1] and Stanley L. Robertson[2]

## ABSTRACT

It is well known that Einstein General Relativity can be expressed covariantly in bi-metric spacetime[1], without the uncertainties which arise from the effects of gravitational energy-momentum pseudo-tensors. However the effect that the Strong Principle of Equivalence (SPOE) has on the underlying curved and flat tetrad bi-metric spacetime structure has not been fully explored. In this paper we show that in the context of bi-metric General Relativity the (SPOE) requirements that:

    **a)** in the absence of gravitation due to spacetime curvature a global Inertial Cartesian Minkowski (ICM) frame exists in which Special Relativity is valid, and,

    **b)** in the presence of gravitation due to spacetime curvature a bi-metric Local Free Fall frame (LFF) exists,

imply the existence of a symmetric gravitational potential tensor $\Phi_\mu{}^\nu = \Phi^\nu{}_\mu$ which is defined thru an exponential connection to the tetrad inner product between the curved symmetric tetrads and the flat symmetric tetrads in the bi-metric spacetime. Hence it follows that, in the bi-metric spacetime with flat metric $\gamma_{\alpha\beta}$ and the curved metric $g_{\mu\nu}$, the (SPOE) requires the curved metric to have the bi-metric Light Cone Causality (LCC) conserving exponential form given by

$$g_{\mu\nu} = [\exp(\Phi)_\nu{}^\alpha] \, \gamma_{\alpha\beta} \, [\exp(\Phi)^\beta{}_\mu].$$

Subject to an appropriate choice of tensor gauge conditions, the substitution of this (LCC) conserving exponential metric into the bi-metric Einstein field equations yields an equivalent set of (LCC) conserving Einstein field equations for the gravitational potential tensor $\Phi_\mu{}^\nu$ the implications of which we further explore in this paper.

It this context the Strong Principle of Equivalence (SPOE) applied to the bi-metric theory of General Relativity implies a new physical paradigm which defines the spacetime metric in an exponential manner using gravitational potential tensors, which allows the operational procedure of *local* spacetime measurements seen by an observer in a *general* spacetime frame of reference to be defined in a similar manner as that of Special Relativity. Hence while an observer in local free fall frame of reference will locally the measure the speed of light as $v = c$, the exponential metric inherent within bi-metric General Relativity implies that an observer in a general bi-metric frame of reference, which contains both the effects of real gravitation due to spacetime curvature and artificial gravitation due to non-inertial coordinate transformations, will also locally measure the vacuum velocity of light to be $v = c$.

[1] FSTC, Charlottesville, VA, USA. E-mail: dleiter@aol.com

[2] Department of Physics, Southwestern Oklahoma State University, Weatherford, OK, USA. E-mail: roberts@swosu.edu

# I  BI-METRIC TETRAD FORMULATION OF GENERAL RELATIVITY AND THE LIGHT CONE CAUSALITY CONSERVING EXPONENTIAL METRIC

The bi-metric description of spacetime (Rosen [1]) a review of which is given in Appendix A assumes that at every point in spacetime a flat background spacetime metric $\gamma_{\mu\nu}$ and a curved physical spacetime metric $g_{\mu\nu}$ are juxtaposed upon each other in a manner so as to allow an operational definition of physical tensor quantities in terms of background covariant derivatives. The bi-metric description of spacetime has the distinct advantage of allowing a well-defined tensor description of gravitational energy-momentum to be defined in the Einstein Theory of General Relativity within a class of gravitational gauge transformations.

In order to see how this occurs we study the effects of the Strong Principle of Equivalence (SPOE) in General Relativity when it is incorporated into the context of the bi-metric description of spacetime. In the following we will show that the bi-metric spacetime the SPOE based assumptions that **a)** in the absence of gravitation due to spacetime curvature a global Inertial Cartesian Minkowski (ICM) frame exists in which Special Relativity is valid, and **b)** in the presence of gravitation due to spacetime curvature a Local Free Fall (LFF) exists which implies the existence of a symmetric exponential gravitational potential tensor connection associated with the tetrad inner product between the curved symmetric tetrads $\omega_{\mu(\alpha)} = \omega_{(\alpha)\mu}$ and the flat symmetric tetrads $\lambda_{\mu(\alpha)} = \lambda_{(\alpha)\mu}$ in the bi-metric spacetime given by

$$\exp(\Phi)_\mu{}^\nu = [\omega_\mu{}^{(\alpha)} \lambda_{(\alpha)}{}^\nu] = [\lambda^\nu{}_{(\alpha)} \omega^{(\alpha)}{}_\mu] = \exp(\Phi)^\nu{}_\mu \qquad (1)$$

which implies that the curved metric is then given by the bi-metric light cone causality conserving exponential form

$$g_{\mu\nu} = [\exp(\Phi)_\nu{}^\alpha] \gamma_{\alpha\beta} [\exp(\Phi)^\beta{}_\mu] \qquad (2)$$

This result then allows us to show in Appendix B, C, and D of this paper how the Strong Principle of Equivalence (SPOE) applied to the bi-metric formulation of General Relativity represents a new physical paradigm which extends the two fundamental rules of observation implied by Special Relativity to include the requirement that the metric of curved spacetime should be defined in an operational manner so that the procedure of local spacetime measurements, as seen by an observer in a general spacetime frame which contains gravitational fields due to spacetime curvature, becomes locally defined in a similar manner as that for the extension of Special Relativity into general flat spacetimes. While it is still true that an observer in a local free fall frame of reference will locally measure the speed of light to be $v = c$. the new paradigm will also requires that an observer in a general frame of reference (which contains both the effects of real gravitation due to spacetime curvature and artificial gravitation due to non-inertial coordinate transformations) will also locally measure the vacuum velocity of light to be v = c. In this context we also show how a well defined General Relativistic N-body formalism can be developed (see Appendix E)

We begin by noting that the bi-metric space-time the global flat space line element is given by

$$d\sigma^2 = \gamma_{\mu\nu} \, dx^\mu \, dx^\nu \qquad (3)$$

and the curved space line element given by

$$ds^2 = g_{\mu\nu} \, dx^\mu \, dx^\nu \qquad (4)$$

In this bi-metric spacetime context there exists flat spacetime tetrads $\lambda_\mu{}^{(\alpha)}$ and curved spacetime tetrads $\omega_\mu{}^{(\alpha)}$ so that the flat and curved background spacetime metric tensors are given respectively by

$$\gamma_{\mu\nu} = \lambda_\mu{}^{(\alpha)} \, \gamma_{(\alpha\beta)} \, \lambda^{(\beta)}{}_\nu \qquad g_{\mu\nu} = \omega_\mu{}^{(\alpha)} \, \gamma_{(\alpha\beta)} \, \omega^{(\beta)}{}_\nu \qquad (5)$$

where the tetrad indices are denoted by indicies in parentheses, and since the flat metric obeys $\gamma_{\mu\nu}{}_{|\rho} = 0$ in the bi-metric spacetime this implies that the flat tetrads obey $\lambda_\mu{}^{(\alpha)}{}_{|\rho} = 0$. In this bi-metric context the existence of a real gravitational field creates a physical difference between the curved spacetime tetrads and the flat background spacetime tetrads. If we assume that curved and flat tetrads in the bi-metric spacetime are symmetric so that $\omega_{\mu(\alpha)} = \omega_{(\alpha)\mu}$ $\lambda_{\mu(\alpha)} = \lambda_{(\alpha)\mu}$ then this implies that the mixed tetrads are symmetric as $\omega_\mu{}^{(\alpha)} = \omega^{(\alpha)}{}_\mu$ and $\lambda_\mu{}^{(\alpha)} = \lambda^{(\alpha)}{}_\mu$ and the physical difference between the two tetrad structures manifests itself thru the existence of a symmetric tensor function $F_\mu{}^\nu = F^\nu{}_\mu$ defined by a bi-metric connection between the curved symmetric tetrads and the flat symmetric tetrads given bu

$$F_\mu{}^\nu = [\omega_\mu{}^{(\alpha)} \, \lambda_{(\alpha)}{}^\nu] = [\lambda^\nu{}_{(\alpha)} \, \omega^{(\alpha)}{}_\mu] = F^\nu{}_\mu \qquad (6)$$

From equations (3) thru (6) it follows immediately that the curved metric tensor $g_{\mu\nu}$ can be written in terms the symmetric tensor $F_\mu{}^\nu$ and the flat background spacetime metric tensor $\gamma_{\mu\nu}$ as

$$g_{\mu\nu} = F_\mu{}^\rho \, \gamma_{\rho\sigma} \, F^\sigma{}_\nu \qquad \gamma_{\rho\sigma} = \lambda_\rho{}^{(\alpha)} \, \gamma_{(\alpha\beta)} \, \lambda^{(\beta)}{}_\sigma \qquad (7)$$

Satisfaction of the (SPOE), in the bi-metric form of Einstein General Relativity, requires that in the presence of gravitation in the Inertial Cartesian Minkowski (ICM) bi-metric spacetime where Special Relativity is valid (i.e. $\gamma = \eta$) access to Local Free Fall (LFF) frames must at every space-time point. In the bi-metric spacetime the LFF condition occurs by locally transforming the bi-metric spacetime coordinates from the global ICM $\{g, \eta\}$ spacetime to a locally non-inertial $\{g', \gamma'\}$ spacetime so that the associated non-inertial forces compensate and cancel out the local gravitational forces. Since for general coordinate transformations $g'_{,\mu} = \gamma'_{,\mu} + g'_{|\mu}$ then local free fall condition $g'_{,\mu} = 0$ will occur if the non-inertial coordinate transformation to the LFF satisfies the condition $\gamma'_{,\mu} = -g'_{|\mu} \neq 0$ *(see Appendix A,B,C for a general review of the properties of bi-metric spacetime, and Appendix D for a more detailed discussion of LFF)*.

Reverting now to matrix notation where { } denote tetrad matricies we note that the LFF implies the condition $\gamma'_{,\mu} = -g'_{|\mu} \neq 0$ with $\{\omega'\} = \{I\}$ and $\{\lambda'\} = F'^{-1}\{I\}$ where $F' \neq 0$. Then LFF corresponds to a transformation to a flat spacetime metric which is is non-inertial with

$$\gamma' = \{\lambda'\}\{\eta\}\{\lambda'\} \neq \eta \qquad (8)$$

while the local curved spacetime metric is given by

$$g' = \{\omega'\}\{\eta\}\{\omega'\} = F' \gamma' F' = \eta \qquad (9)$$

Since $\{\omega'\} = \{I\}$ in the LFF frame then this implies that

$$\{\omega'\}_{|\mu} \{\omega'\}^{-1} = \{\omega'\}^{-1} \{\omega'\}_{|\mu} \qquad (10)$$

However since this is a tensor relationship it follows that

$$\{\omega\}_{|\mu} \{\omega\}^{-1} = \{\omega\}^{-1} \{\omega\}_{|\mu} \qquad (11)$$

is valid in all frames. However since $\{\omega\} = F \{\lambda\}$ and $\{\lambda\}_{|\mu} = 0$ then this implies that in all frames that

$$[F_{|\mu} \{\lambda\}, F^{-1}\{\lambda\}] = 0 \qquad (12)$$

Now going to the global ICM frame where $\{\lambda\} = \{I\}$ this implies that

$$[F_{|\mu}, F^{-1}] = 0 \qquad (13)$$

which being a tensor equation is also true in all frames. This implies that the Strong Principle of Equivalence requirement that an LFF exists everywhere in the bi-metric spacetime implies that a symmetric gravitational potential tensor function $\Phi = \Phi_\mu{}^\nu = \Phi^T = \Phi^\mu{}_\nu$ exists everywhere in the bi-metric spacetime by the relationship $\Phi_{|\mu} = F_{|\mu} F^{-1} = F^{-1} F_{|\mu}$. Hence we can solve this relationship to obtain the connection between the symmetric tensor function F and the symmetric gravitational potential tensor function $\Phi$ as $F = \exp(\Phi)$. Since the above equations also imply that

$$[\exp(\Phi)_{|\mu}, \exp(-\Phi)] = 0 \qquad (14)$$

this implies that in the bi-metric spacetime the gravitational potential tensor must also obey

$$[\Phi_{|\mu}, \Phi] = 0 \qquad (15)$$

In summary the requirement of the existence of the LFF implies that in general

$$\{\omega\} = \exp(\Phi)\{\lambda\} = \{\omega\}^T \qquad (16\text{-a})$$

$$g = \omega\}\{\eta\}\{\omega\} = \exp(\Phi)\,\gamma\,\exp(\Phi) \qquad (16\text{-b})$$

$$\gamma = \{\lambda\}\,\{\eta\}\,\{\lambda\} \qquad (16\text{-c})$$

where in the above $\{\lambda\} = \lambda_\mu{}^{(\alpha)}$ and $\{\eta\} = \eta_{(\alpha\beta)}$.

In the context of the global Inertial Cartesian Minkowski spacetime (ICM) where $\gamma = \eta$ is diagonal in the presence of gravitation, the tetrad quantities are given by

$$\{\lambda\} = \{I\} \quad , \quad \{\omega\} = \exp(\Phi)\,\{I\} \qquad (17\text{-a})$$

with the metric tensors

$$\gamma = \eta \quad , \quad g = \exp(\Phi)\,\eta\,\exp(\Phi) \qquad (17\text{-b})$$

From this it follows that

$$g_{|\mu}\, g^{-1} = \Phi_{|\mu} + g\,\Phi_{|\mu}\,g^{-1} \qquad (17\text{-c})$$

However since we proved earlier that $[\Phi_{|\mu}, \Phi] = 0$ it follows that

$$\Phi_{|\mu} = 1/2 \, g_{|\mu} \, g^{-1} \tag{17-d}$$

which when written out in its full spacetime index form is given by

$$\Phi_\alpha{}^\beta{}_{|\mu} = 1/2 \, g_{\alpha\rho|\mu} \, g^{\rho\beta} \tag{17-e}$$

Now in bi-metric spacetime the curve metric tensor density is defined as $\mathbf{g}_{\mu\nu} = (-\kappa)^{-1/2} g_{\mu\nu}$ where $(-\kappa)^{1/2} = (-g)^{1/2}/(-\gamma)^{1/2} = |\exp(\Phi)|$. In order to calculate the general form of $(-\kappa)^{1/2}$ we first note that since $\Phi$ is a symmetric 4x4 spacetime matrix it can be locally diagonalized by a local coordinate transformation so that locally the relationship $|\exp(\Phi')| = \exp[\text{Tr}(\Phi')]$ holds. However since this is a tensor relationship it must be true in a general frame of reference and hence it follows that

$$(-\kappa)^{1/2} = (-g)^{1/2}/(-\gamma)^{1/2} = |\exp(\Phi)| = \exp[\text{Tr}(\Phi)] \tag{18}$$

must be true in all frames in the bi-metric spacetime.

*Using the above result it follows that the (SPOE) implies that curved metric tensor density is given by*

$$\mathbf{g} = (-\kappa)^{-1/2} g = \exp(-2\phi) \, \mathbf{g} \, \exp(-2\phi) \tag{19}$$

*where* $\phi = 1/4 \, [\text{Tr}(\Phi) \, I - 2\Phi]$ *or upon inverting* $\Phi = \text{Tr}(\phi) \, I - 2\phi]$.

Hence for the metric tensor density $\mathbf{g} = (-\kappa)^{-1/2} g \quad \mathbf{g}^{-1} = (-\kappa)^{1/2} g^{-1}$

$$\mathbf{g}_{|\mu} = -2(\phi_{|\mu} \mathbf{g} + \mathbf{g} \, \phi_{|\mu}) \tag{20-a}$$

$$-1/2 \, \mathbf{g}_{|\mu} \, \mathbf{g}^{-1} = \phi_{|\mu} + \mathbf{g} \, \phi_{|\mu} \, \mathbf{g}^{-1} \tag{20-b}$$

and using the fact proved earlier that $[\phi_{|\mu}, \phi] = 0$ we have

$$\phi_{|\mu} = -1/4 \, \mathbf{g}_{|\mu} \, \mathbf{g}^{-1} \qquad (21\text{-a})$$

or in full spacetime index form

$$\phi_\alpha{}^\beta{}_{|\mu} = -1/4 \, \mathbf{g}_{\alpha\rho|\mu} \, \mathbf{g}^{\rho\beta} \qquad (21\text{-b})$$

In the bi-metric spacetime if we insert the connection

$$\phi_\mu{}^\nu{}_{|\alpha} = -1/4 \, (\mathbf{g}_{\mu\lambda|\alpha} \, \mathbf{g}^{\lambda\nu}) \qquad (21\text{-c})$$

into the Einstein tensor density $\mathbf{G}_\mu{}^\nu$ in the Einstein equation

$$-1/2 \, \mathbf{G}_\mu{}^\nu = (4\pi G / c^2) \, \mathbf{t}_\mu{}^\nu \qquad (22\text{-a})$$

this converts these equations into non-linear bi-metric tensor density equations for the gravitational potential tensor $\phi_\mu{}^\nu$ given by

$$\{[(-\kappa)^{1/2}(\phi_\mu{}^{\nu|\alpha} - \phi_\mu{}^{\alpha|\nu})]$$
$$+ \delta_\mu{}^\nu (\mathbf{g}^{\alpha\rho} \phi_\rho{}^\beta{}_{|\beta}) - \delta_\mu{}^\alpha (\mathbf{g}^{\nu\rho} \phi_\rho{}^\beta{}_{|\beta})\}_{|\alpha}$$
$$= [(4\pi G / c^2) \, \mathbf{t}_\mu{}^\nu + \mathbf{U}_\mu{}^\nu] \qquad (22\text{-b})$$

where $(-\kappa)^{1/2} = [(-g)^{1/2} / (-\gamma)^{1/2}] = \exp[2 \, \text{Tr}(\phi)]$ and $\mathbf{U}_\mu{}^\nu$ is the Einstein gravitational energy–momentum tensor density

$$\mathbf{U}_\mu{}^\nu = (-\kappa)^{1/2} [1/2 \, \delta_\mu{}^\nu W_\alpha{}^\alpha - W_\mu{}^\nu] \qquad (22\text{-c})$$

$$W_\mu{}^\nu = 2\,[(\phi_\rho{}^\sigma{}_{|\mu} \phi_\sigma{}^{\rho|\nu}) - 1/2 \, \delta_\mu{}^\nu \, \text{Tr}(\phi_\rho{}^\sigma{}_{|\lambda} \phi_\sigma{}^{\rho|\lambda})\,]$$

$$+ \,[(\phi_\rho{}^\rho)_{|\mu} (\phi_\sigma{}^\sigma)^{|\nu} - 1/2 \, \delta_\mu{}^\nu (\phi_\rho{}^\rho)^{|\lambda} (\phi_\sigma{}^\sigma)_{|\lambda} \qquad (22\text{-d})$$

$$+ \,[g^{\nu\rho} (\phi_\rho{}^\lambda{}_{|\mu}) (\phi_\lambda{}^\alpha{}_{|\alpha})\}$$

We can simplify the above bi-metric gravitational potential form of the Einstein equations by choosing the Harmonic Tensor Gauge condition $\phi_\mu{}^\nu{}_{|\nu} = 0$ which is equivalent to $\mathbf{g}^{\mu\nu}{}_{|\nu} = 0$. Then the bi-metric tensor form of the Einstein equation becomes simplifies to become

$$\{[\exp(2\mathrm{Tr}(\phi))] \, [(\phi_\mu{}^{\nu \, | \, \alpha} - \phi_\mu{}^{\nu | \, \nu})]\}_{|\alpha} = [\exp(2\mathrm{Tr}(\phi))] \, [(4\pi G/c^2)\,\tau_\mu{}^\nu + \mathbf{U}_\mu{}^\nu] \qquad (23\text{-a})$$

where the Einstein Gravitational energy momentum tensor is given by

$$\mathbf{U}_\mu{}^\nu = \{\delta_\mu{}^\nu [\mathrm{Tr}\,(\phi^{\,|\,\alpha})\,\mathrm{Tr}\,(\phi_{\,|\,\alpha})] - \mathrm{Tr}\,(\phi_{\,|\,\mu})\,(\phi^{T\,|\,\nu})\} \qquad (23\text{-b})$$

In the bi-metric form of Einstein equation the bi-metric the issue of ***Bi-metric Light Cone Causality*** [2,3,4] revolves around the question as to whether timelike worldines associated with $ds^2 > 0$ in the context of the curved spacetime will always imply that $d\sigma^2 > 0$ will also hold for these worldline so that they are also timelike in the context of the flat background spacetime. However, since the exponential form of the bi-metric g-metric tensors,

$$g(\Phi)_{\mu\nu} = [\exp(\Phi)_\mu{}^\rho]\,\gamma_{\rho\sigma}\,[\exp(\Phi)^\sigma{}_\nu]\,, \qquad \mathbf{g}(\phi)_{\mu\nu} = [\exp(-2\phi)_\mu{}^\rho]\,\mathbf{g}_{\rho\sigma}\,[\exp(-2\phi)^\sigma{}_\nu] \qquad (24\text{-a})$$

$$g(-\Phi)^{\mu\nu} = [\exp(-\Phi)^\mu{}_\rho]\,\gamma^{\rho\sigma}\,[\exp(-\Phi)_\sigma{}^\nu]\,, \qquad \mathbf{g}(-\phi)^{\mu\nu} = [\exp(2\phi)^\mu{}_\rho]\,\mathbf{g}^{\rho\sigma}\,[\exp(2\phi)_\sigma{}^\nu] \qquad (24\text{-b})$$

with $\Phi_\mu{}^\nu = [\mathrm{Tr}\,(\phi)\,\delta_\mu{}^\nu - 2\phi_\mu{}^\nu]$ and $\phi_\mu{}^\nu = 1/4\,[\mathrm{Tr}\,(\Phi)\,\delta_\mu{}^\nu - 2\Phi_\mu{}^\nu]$ were derived under the Strong Principle of Equivalence (SPOE) requirement that local free fall frames (LFF) exist everywhere in the bi-metric $\{g, \gamma\}$ spacetime, then we would expect that these exponential g-metric tensors should dynamically preserve the ***Bi-metric Light Cone Causality*** connection between the light cones in the curved physical and flat background spacetimes respectively in the context of the solutions to the above bi-metric Einstein equation. We will demonstrate this in the ICM bi-metric spacetime *where in all frames of reference and for all values of the gravitational potential tensor* $\Phi$ *timelike world lines in the curved spacetime obey*

$$ds^2(\Phi) = g(\Phi)_{\mu\nu}\,dx^\mu\,dx^\nu = [\exp(\Phi)_\mu{}^\alpha]\,\eta_{\alpha\beta}\,[\exp(\Phi)^\beta{}_\nu]\,dx^\mu\,dx^\nu > 0 \qquad (25)$$

where the underlying flat spacetime element is given by $d\sigma^2 = \eta_{\mu\nu}\,dx^\mu\,dx^\nu$

Now since $|\exp(\Phi)| = \exp[\text{Tr}(\Phi)] > 0$ then the symmetric spacetime matrix tensor quantity $\exp(\Phi)_\mu{}^\nu$ always has an inverse given by $\exp(-\Phi)_\mu{}^\nu$ and hence we can always define the coordinate transformation

$$dx'^\mu = \exp(\Phi)^\mu{}_\alpha \, dx^\alpha \quad \text{and its inverse} \quad dx^\mu = \exp(-\Phi)^\mu{}_\alpha \, dx'^\alpha \tag{26-a}$$

$$dx'_\mu = \exp(\Phi)_\mu{}^\alpha \, dx_\alpha \quad \text{and its inverse} \quad dx_\mu = \exp(-\Phi)_\mu{}^\alpha \, dx'_\alpha \tag{26-b}$$

and suppose that under this coordinate transformation where

$$(\Phi')_\mu{}^\nu = [\exp(\Phi)_\beta{}^\nu \exp(-\Phi)_\mu{}^\alpha] \, \Phi_\alpha{}^\beta \tag{26-c}$$

$$\exp(\Phi')_\mu{}^\nu = [\exp(\Phi)_\beta{}^\nu \exp(-\Phi)_\mu{}^\alpha] \exp(\Phi)_\alpha{}^\beta = \exp(\Phi)_\beta{}^\nu \, \delta_\mu{}^\beta = \exp(\Phi)_\mu{}^\nu \tag{26-d}$$

$$d\sigma^2 = d\sigma'^2 = \gamma'_{\mu\nu} \, dx'^\mu \, dx'^\nu \tag{26-e}$$

we have that

$$\begin{aligned} ds^2 = ds'^2 &= g'(\Phi')_{\mu\nu} \, dx'^\mu dx'^\nu \quad > 0 \\ &= [\exp(\Phi')_\mu{}^\alpha] \, \gamma'_{\alpha\beta} \, [\exp(\Phi')_\beta{}^\nu] \, dx'^\mu dx'^\nu \\ &= \eta_{\mu\nu} \, dx'^\mu dx'^\nu \end{aligned} \tag{26-f}$$

Then this implies that

$$g'(\Phi')_{\mu\nu} = [\exp(-\Phi)_\mu{}^\beta] \, g_{\alpha\beta} \, [\exp(-\Phi)^\beta{}_\nu] = [\exp(\Phi')_\mu{}^\alpha] \, \gamma'_{\alpha\beta} \, [\exp(\Phi')^\beta{}_\nu] = \eta_{\mu\nu} \tag{26-g}$$

$$\gamma'_{\mu\nu} = [\exp(-\Phi)_\mu{}^\beta] \, \eta_{\alpha\beta} \, [\exp(-\Phi)^\beta{}_\nu] = \{\exp(-\Phi')_\mu{}^\alpha \, \eta_{\alpha\beta} \, \exp(-\Phi')^\beta{}_\nu\} \tag{26-h}$$

Now r since $|\exp(\Phi')| = |\exp(\Phi)| = \exp[\text{Tr}(\Phi)] > 0$ then the spacetime matrix tensor $\exp(\Phi')$ always has an inverse given by $\exp(-\Phi')$. This implies that a non-inertial local free fall (LFF) frame exists (see Appendix D) *such that locally at the free fall space time point* $x = x_0$ *the local gravitational forces are locally canceled out by non-inertial forces and*

$$g'' = g' \,, \quad \Phi'' = \Phi' \,, \quad \gamma'' = \gamma' \,, \quad g''_{,\mu} = 0 \,, \quad \gamma''_{,\mu} = -g''|_\mu \tag{27-a}$$

and the local curved metric is Minkowskian

$$g'' = g' = [\exp(\Phi')] \, \gamma' \, [\exp(\Phi')] = \eta \qquad (27\text{-b})$$

which requires that the background spacetime metric locally transform as

$$\gamma'' = [\exp(-\Phi'')] \, \eta \, [\exp(-\Phi'')] = g''(-\Phi'') \neq g''(\Phi) \qquad (27\text{-c})$$

This implies that in the context of LFF at a spacetime point

$$ds''(\Phi'')^2 = \eta_{\mu\nu} \, dx''^{\mu} \, dx''^{\nu} = ds'(\Phi')^2 = ds(\Phi)^2 > 0 \qquad (27\text{-d})$$

$$d\sigma''^2 = \gamma''_{\mu\nu} \, dx''^{\mu} \, dx''^{\nu} = g''(-\Phi'')_{\mu\nu} \, dx''^{\mu} \, dx''^{\nu} = ds''^2(-\Phi'') > 0 \qquad (27\text{-e})$$

since $ds''^2(\Phi'') > 0$ was assumed to be true in all frames of reference and for all values of $\Phi''$.

since this is a tensor equation then transforming back from the LFF to the general frame it follows that

$$d\sigma^2 = d\sigma''^2 = ds''^2(-\Phi'') > 0 \qquad (25\text{-f})$$

*which proves that LCC holds in the bi-metric spacetime if the metric has the exponential*

*form given by (24) and (25). This is consistent because the exponential form of the metric*

*implies that*

$$d\sigma^2 = d\sigma''^2 = ds''^2(-\Phi'') \neq ds''^2(\Phi'') = ds^2(\Phi) \qquad (27\text{-g})$$

*Hence even though* $d\sigma^2 > 0$ *this does not imply that it is equal to* $ds^2(\Phi) > 0$.

## III. CONCLUSIONS

We have shown that in the context of bi-metric General Relativity the (SPOE) requirements that:

   **a)** in the absence of gravitation due to spacetime curvature a global Inertial Cartesian Minkowski (ICM) frame exists in which Special Relativity is valid, and,

   **b)** in the presence of gravitation due to spacetime curvature a bi metric Local Free Fall frame (LFF) exists,

imply the existence of a symmetric gravitational potential tensor $\Phi_\mu{}^\nu = \Phi^\nu{}_\mu$ which is defined thru an exponential connection to the tetrad inner product between the curved symmetric tetrads and the flat symmetric tetrads in the bi-metric spacetime. It follows that in the the (SPOE) requires that, in the bi-metric spacetime with flat metric $\gamma_{\alpha\beta}$ and the curved metric $g_{\mu\nu}$ must have a bi-metric Light Cone Causality (LCC) conserving exponential form given by

$$g_{\mu\nu} = [\exp(\Phi)_\nu{}^\alpha] \; \gamma_{\alpha\beta} \; [\exp(\Phi)^\beta{}_\mu]$$

Subject to an appropriate choice of tensor gauge conditions, the substitution of this (LCC) conserving exponential metric into the bi-metric Einstein field equations yields an equivalent set of (LCC) conserving Einstein field equations for the gravitational potential tensor $\Phi_\mu{}^\nu$ *which allowed a fully nonlinear N-body approximation to be developed.*

It conclusion we note that the Strong Principle of Equivalence (SPOE) applied to the bi-metric theory of General Relativity[1] implies a new physical paradigm which defines the spacetime metric in an exponential manner using gravitational potential tensors, which allows the operational procedure of *local* spacetime measurements seen by an observer in a *general* spacetime frame of reference to be defined in a similar manner as that of Special Relativity.

Hence while an observer in local free fall frame of reference will still locally the measure the local speed of light as v = c, bi-metric Einstein General Relativity also implies that an observer in a general frame of reference, which contains both the effects of real gravitation due to spacetime curvature and artificial gravitation due to non-inertial coordinate transformations, will also locally measure the vacuum velocity of light to be v = c.

## REFERENCES

[1] Rosen, N., (1963), Annals of Physics, 22, pg 1.

[2] Pitts, J.B., and Schieve, W.C., (2001), General Relativity and Gravitation, 33, pg. 1319

[3] Pitts, J.B., (2001), "Null Cones in Lorentz-Covariant General Relativity" gr-qc/0111004

[4] Pitts, J.B. and Schieve, W.C., "Light Cone Consistency in Bimetric General Relativity" gr-qc/010197

APPENDIX A: THE RELATIVISTIC BI-METRIC DESCRIPTION OF CURVED SPACETIME

In order to operationally define the dynamic effects of gravitational energy momentum in curved spacetime, independent of the flat geometric and inertial properties of the curved spacetime within which the effects of gravitation act, we assume (Rosen (1963) [1] ) that spacetime is fundamentally bi-metric in nature.

This means that there exists at every point of spacetime a flat relativistic background metric tensor $\gamma_{\mu\nu}$ for which the Riemann-Christoffel curvature tensor $r_{\mu\nu\alpha\beta}$ is assumed to vanish everywhere identically. The relativistic flat background metric tensor $\gamma_{\mu\nu}$ exists in addition to that of the curved metric tensor $g_{\mu\nu}$ for which the Riemann-Christoffel curvature tensor $R_{\mu\nu\alpha\beta}$ is assumed to be non-zero. The physical metric interval $ds^2$ and relativistic background metric interval $ds^2$ in spacetime are given by

$$ds^2 = g_{\mu\nu} dx^\mu dx^\nu \qquad d\sigma^2 = \gamma_{\mu\nu} dx^\mu dx^\nu \qquad \text{(A.1-a)}$$

Since the $\gamma_{\mu\nu}$ describes flat background spacetime, the most general change that $\gamma_{\mu\nu}$ can undergo corresponds to an arbitrary flat spacetime coordinate transformation and which involves four arbitrary tensor functions. Hence in this context it follows that without loss of generality that we are allowed to impose four additional tensor conditions on the metrics $g_{\mu\nu}$ and the $\gamma_{\mu\nu}$ in order to fix the relationship between them. However since these four additional metric connection conditions are *tensor conditions* they are more general than the usual they do not single out a specific coordinate system.

In the absence of real gravitational forces due to spacetime curvature $g_{\mu\nu}$ is equal to $\gamma_{\mu\nu}$. In the presence of real gravitational forces due to spacetime curvature $g_{\mu\nu}$ is not equal to $\gamma_{\mu\nu}$. In this context, the difference between the physical metric $g_{\mu\nu}$ and its first and second derivatives from the flat background spacetime metric $\gamma_{\mu\nu}$ and its first and second derivatives represents operational data by which the geometry of a given curved spacetime with gravitation may be compared with that of the geometry of the spacetime one would have if the gravitational field were removed.

*In this context the initial choice of the flat background spacetime metric $\gamma_{\mu\nu}$ must be made on the basis of physical considerations.* This can be accomplished by using the fact that in the absence of real gravitational forces due to spacetime curvature the Cartesian coordinate frame of Special Relativity represents the global definition of "inertial frame" for which the non-gravitational laws of physics are covariant under the constant velocity Lorentz transformations associated with the Minkowski spacetime metric $\eta_{\mu\nu}$.

Specifically this means that in the ***absence*** of real gravitational forces due to spacetime curvature, physical spacetime must be the inertial Cartesian Minkowski spacetime of Special Relativity. In this case this implies that $\gamma_{\mu\nu} = \eta_{\mu\nu}$, $g_{\mu\nu} = \eta_{\mu\nu}$ and

$$ds^2 = \eta_{\mu\nu} dx^\mu dx^\nu \qquad d\sigma^2 = \eta_{\mu\nu} dx^\mu dx^\nu \qquad (A.2\text{-a})$$

In the ***presence*** of real gravitation due to curvature, in the physical inertial Cartesian Minkowski spacetime of Special Relativity, $g_{\mu\nu}$ is different from $\eta_{\mu\nu}$ and

$$ds^2 = g_{\mu\nu} dx^\mu dx^\nu \qquad d\sigma^2 = \eta_{\mu\nu} dx^\mu dx^\nu \qquad (A.2\text{-b})$$

However when we make general coordinate transformations $dx'^\mu = (\partial x'^\mu / \partial x^\nu) dx'^\nu$ from the physical inertial Cartesian Minkowski spacetime of Special Relativity $dx^\mu$ to a non-inertial and or non-cartesian spacetime $dx'^\mu$ these coordinate transformations act on *both* the physical spacetime (g-metric) *and* the inertial flat background spacetime ($\eta$ -metric) *together* giving

$$g'_{\mu\nu} = (\partial x^\rho / \partial x^{\mu'}) (\partial x^\sigma / \partial x^{\nu'}) g_{\rho\sigma} \qquad \gamma'_{\mu\nu} = (\partial x^\rho / \partial x^{\mu'}) (\partial x^\sigma / \partial x^{\nu'}) \eta_{\rho\sigma} \qquad (A.3\text{-a})$$

$$ds'^2 = g'_{\mu\nu} dx'^\mu dx'^\nu \qquad d\sigma'^2 = \gamma'_{\mu\nu} dx'^\mu dx'^\nu \qquad (A.3\text{-b})$$

Note that the inertial background spacetime "$\eta$-metric" transforms in the same relativistic tensor manner as that of the "g-metric" under general coordinate transformations. This choice of Cartesian Minkowski spacetime of Special Relativity as the global definition of inertial frame causes the invariant scalar quantity $(-\kappa)^{1/2} = (-g)^{1/2} / (-\gamma)^{1/2}$ to have the value

$$(-\kappa')^{1/2} = (-g')^{1/2} / (-\gamma')^{1/2} = (-\kappa)^{1/2} = (-g)^{1/2} / (-\eta)^{1/2} = (-g)^{1/2} \qquad (A.4)$$

Now in the usual manner associated with the g-metric covariant differentiation (denoted by " ; " ) there will be a "non-tensor g- Christoffel 3-index symbol" $\Gamma^\mu_{\nu\alpha}$ associated with $g_{\mu\nu\,;\,\lambda} = 0$ given by

$$\Gamma^\lambda_{\mu\nu} = 1/2 \; g^{\lambda\alpha} (\partial_\nu g_{\mu\alpha} + \partial_\mu g_{\nu\alpha} - \partial_\alpha \gamma_{\mu\nu}) \qquad (A.5)$$

and in a similar manner associated with the γ-metric covariant differentiation

(denoted by " | " ) there will be a "non-tensor γ - Christoffel 3-index symbol" $Z^\lambda{}_{\mu\nu}$

associated with $\gamma_{\mu\nu} | \lambda = 0$ given by

$$Z^\lambda{}_{\mu\nu} = 1/2 \, \gamma^{\lambda\alpha} (\partial_\nu \gamma_{\mu\alpha} + \partial_\mu \gamma_{\nu\alpha} - \partial_\alpha \gamma_{\mu\nu}) \tag{A.6}$$

*Now under general spacetime co-ordinate transformations denoted by*

$$dx'^\mu = (\partial x'^\mu / \partial x^\nu) \, dx^\nu \tag{A.7-a}$$

*both non-tensor Christoffel-3-index symbols* $\Gamma^\lambda{}_{\mu\nu}$ *and* $Z^\lambda{}_{\mu\nu}$ *transform in the same*

*non-covariant non-tensor like coordinate frame dependent manner as*

$$\Gamma'^\lambda{}_{\mu\nu} = (\partial x^\sigma / \partial x'^\mu)(\partial x^\tau / \partial x'^\mu)(\partial x'^\lambda / \partial x^\rho) \, \Gamma^\rho{}_{\sigma\tau}$$
$$+ (\partial^2 x^\sigma / \partial x'^\mu \partial x'^\nu)(\partial x'^\lambda / \partial x^\sigma) \tag{A.7-b}$$

$$Z'^\lambda{}_{\mu\nu} = (\partial x^\sigma / \partial x'^\mu)(\partial x^\tau / \partial x'^\mu)(\partial x'^\lambda / \partial x^\rho) \, Z^\rho{}_{\sigma\tau}$$
$$+ (\partial^2 x^\sigma / \partial x'^\mu \partial x'^\nu)(\partial x'^\lambda / \partial x^\sigma) \tag{A.7-c}$$

*However the difference between the two non-tensor Christoffel symbols* $\Delta'^\mu{}_{\nu\alpha}$

$$\Delta'^\mu{}_{\nu\alpha} = (\Gamma'^\mu{}_{\nu\alpha} - Z'^\mu{}_{\nu\alpha}) \tag{A.7-a}$$

*will transform covariantly like a tensor as*

$$\Delta'^\lambda{}_{\mu\nu} = (\partial x^\sigma / \partial x'^\mu)(\partial x^\tau / \partial x'^\nu)(\partial x'^\lambda / \partial x^\rho) \, \Delta^\rho{}_{\sigma\tau} \tag{A.7-b}$$

Now let the coordinate transformation

$$\gamma'_{\mu\nu} = (\partial x^\rho / \partial x'^\mu)(\partial x^\sigma / \partial x'^\nu)\eta_{\rho\sigma} \qquad (A.8\text{-a})$$

$$g'_{\mu\nu} = (\partial x^\rho / \partial x'^\mu)(\partial x^\sigma / \partial x'^\nu) g_{\rho\sigma}$$

represent a transformation from a curved spacetime, with $\Gamma^\mu{}_{\nu\alpha} = \Delta^\mu{}_{\nu\alpha} \neq 0$

(associated with a flat background Minkowski spacetime where $Z^\mu{}_{\nu\alpha} = 0$) to another

curved spacetime with $\Gamma'^\mu{}_{\nu\alpha} \neq 0$ (associated with a general flat background

spacetime where $Z'^\mu{}_{\nu\alpha} \neq 0$). Then it follows from the non-tensor transformations

(A.7-b and c) that

$$\Gamma'^\lambda{}_{\mu\nu} = (\partial x^\sigma / \partial x'^\mu)(\partial x^\tau / \partial x'^\mu)(\partial x'^\lambda / \partial x^\rho)\Delta^\rho{}_{\sigma\tau} + Z'^\lambda{}_{\mu\nu} \qquad (A.8\text{-b})$$

where $Z'^\lambda{}_{\mu\nu}$ is a coordinate frame dependent non-tensor quantity given by

$$Z'^\lambda{}_{\mu\nu} = 1/2\, \gamma'^{\lambda\alpha}(\partial_\nu \gamma'_{\mu\alpha} + \partial_\mu \gamma'_{\nu\alpha} - \partial_\alpha \gamma'_{\mu\nu})$$

$$= (\partial^2 x^\sigma / \partial x'^\mu \partial x'^\nu)(\partial x'^\lambda / \partial x^\sigma) \qquad (A.8\text{-c})$$

(This relationship implies that the primed $\Gamma'^\lambda{}_{\mu\nu}$, $\Delta'^\lambda{}_{\mu\nu}$ and $Z'^\lambda{}_{\mu\nu}$ indices can be raised and lowered by the g'-metric in the general background case since the unprimed $\Gamma^\lambda{}_{\mu\nu}$ indices are raised and lowered by the g-metric in the Minkowski background case).

Hence in general from (A.7) and (A.8) we see that the non-tensor "g- Christoffel 3-index symbol $\Gamma'^{\mu}{}_{\nu\alpha}$ which represents the total gravitational force can be formally written as

$$\Gamma'^{\mu}{}_{\nu\alpha} = \Delta'^{\mu}{}_{\nu\alpha} + Z'^{\mu}{}_{\nu\alpha} \qquad (A.9)$$

Hence equation (A.9) represents an explicit breakup of the total gravitational geometry and force carried by the non-tensor Christoffel symbol $\Gamma'^{\mu}{}_{\nu\alpha}$ into an "actual tensor part" $\Delta'^{\mu}{}_{\nu\alpha}$ and a "fictitious coordinate frame dependent non-covariant, non-tensor part" $Z'^{\mu}{}_{\nu\alpha}$ which are physically interpreted as follows:

**a)** the actual gravitational geometry-force carried by the covariant Christoffel tensor $\Delta'^{\mu}{}_{\nu\alpha}$ due to the presence of actual mass-energy distributions, which generate actual curvature in the spacetime, **.b)** the fictitious coordinate frame dependent non-tensor gravitational geometry-force carried by the non-covariant Christoffel non-tensor $Z'^{\mu}{}_{\nu\alpha}$ due to general co-ordinate transformations (which do not generate spacetime curvature) from the flat Spacetime of Special Relativity to general flat non-Cartesian, non-inertial spacetimes.

Hence the *total non-tensor* gravitational geometry and force carried by the Christoffel non-tensor $\Gamma'^{\mu}{}_{\nu\alpha}$ is the physical sum of the covariant *actual tensor* gravitational geometry and force $\Delta'^{\mu}_{\nu\alpha}$ and the non-covariant *fictitious non-tensor* gravitational geometry and force $Z'^{\mu}_{\nu\alpha}$.

This result can be thought of as a manifestation of the Weak Principle of Equivalence. Since the Riemann Christoffel curvature tensor vanishes in general for the arbitrary coordinate transformations (which convert $\eta_{\mu\nu}$ from its inertial Cartesian Minkowski metric form into a non-Cartesian, non-inertial $\gamma_{\mu\nu}$ metric form) it follows that covariant differentiation with respect to the background metric (which we will now call by the name $\gamma$-differentiation and denote by the subscript bar-symbol " | ") can be defined in the same manner as covariant differentiation with respect to the gravitational metric (which we now call g-differentiation and denote by the subscript semicolon symbol " ; ").

However we note that the vanishing curvature in flat $\gamma$-spacetime means that one can interchange the order of "$\gamma$-differentiation" so that it obeys all the rules of ordinary differentiation except that the covariant "$\gamma$-derivative" of the $\gamma_{\mu\nu}$ must vanish.

Dropping the primed notation in what follows and for simplicity now using the shorthand notation $\partial_\mu = $ " , " we now have from the vanishing of the g-covariant derivative $g_{\mu\nu\,;\,\lambda}$ and equation (A.9) that

$$g_{\mu\nu\,;\,\lambda} = (g_{\mu\nu\,,\,\lambda} - \Gamma^{\alpha}{}_{\mu\lambda}\, g_{\alpha\nu} - \Gamma^{\alpha}{}_{\nu\lambda}\, g_{\alpha\mu}) = 0 \qquad \text{(A.10-a)}$$

$$[(g_{\mu\nu\,,\,\lambda} - Z^{\alpha}{}_{\mu\lambda}\, g_{\alpha\nu} - Z^{\alpha}{}_{\nu\lambda}\, g_{\alpha\mu}) - \Delta^{\alpha}{}_{\mu\lambda}\, g_{\alpha\nu} - \Delta^{\alpha}{}_{\nu\lambda}\, g_{\alpha\mu}] = 0 \qquad \text{(A.10-b)}$$

Now using the definition of covariant $\gamma$-differentiation of $g_{\mu\nu}$ given by

$$g_{\mu\nu}\,|\,\lambda = g_{\mu\nu\,,\,\lambda} - Z^{\alpha}{}_{\mu\lambda}\, g_{\alpha\nu} - Z^{\alpha}{}_{\nu\lambda}\, g_{\alpha\mu} \qquad \text{(A.11)}$$

in (A.10) we can write the vanishing of $g_{\mu\nu\,;\,\lambda}$ in the following form

$$g_{\mu\nu\,;\,\lambda} = [(g_{\mu\nu}\,|\,\lambda) - \Delta^{\alpha}{}_{\mu\lambda}\, g_{\alpha\nu} - \Delta^{\alpha}{}_{\nu\lambda}\, g_{\alpha\mu}] = 0 \qquad \text{(A.12)}$$

It then follows directly from the cyclic properties of the indices in (A.12) that the Christoffel 3-index tensor $\Delta^{\mu}{}_{\nu\alpha}$ defined in (A.9) (which is associated with the geometrical effects of pure gravitation acting in the spacetime independent of the geometrical effects of the background metric) can be written as

$$\Delta^{\lambda}{}_{\mu\nu} = 1/2\ g^{\lambda\alpha}\,(g_{\mu\alpha}\,|\,\nu + g_{\nu\alpha}\,|\,\mu - g_{\mu\nu}\,|\,\alpha) \qquad \text{(A.13)}$$

Hence from (A.11 it follows that the non-tensor $g_{\mu\alpha\,,\,\nu}$ breaks up into two parts as

$$g_{\mu\alpha\,,\,\nu} = (g_{\mu\alpha}\,|\,\nu + z_{\mu\alpha\nu}) \qquad \text{(A.14-a)}$$

where

$$z_{\mu\alpha\nu} = (Z^{\alpha}{}_{\mu\lambda}\, g_{\alpha\nu} + Z^{\alpha}{}_{\nu\lambda}\, g_{\alpha\mu}) \qquad \text{(A.14-b)}$$

Equation (A.14) represents a simple expression of the explicit physical breakup of the total gravitational geometry-force carried by the non-tensor spacetime derivative of the metric $g_{\mu\alpha\,,\,\nu}$ into an actual tensor part $g_{\mu\alpha}\,|\,\nu$ and a fictitious non-tensor coordinate frame dependent part given by $z_{\mu\alpha\nu}$.

Substituting (A.9) into the Riemann-Christoffel curvature tensor $R_{\mu\nu}$ and using (A.5) we have ( since $r_{\mu\nu\alpha\beta} = 0$) that

$$R_{\mu\nu} = (-\Delta^{\alpha}{}_{\mu\nu}|_{\alpha} + \Delta^{\alpha}{}_{\alpha\mu}|_{\nu} - \Delta^{\alpha}{}_{\alpha\beta}\Delta^{\beta}{}_{\mu\nu} + \Delta^{\alpha}{}_{\beta\mu}\Delta^{\beta}{}_{\alpha\nu}) \qquad (A.15)$$

This is the curvature tensor $R_{\mu\nu}$ associated with the curvature effects of pure gravitation acting in the spacetime independent of the geometrical effects of the background metric. It has no $\gamma$-metric coordinate transformation "z-dependence" in it since the $\gamma$-metric has vanishing curvature.

In summary we have shown that using the flat background description of spacetime automatically decomposes the ordinary spacetime non-tensors $\Gamma^{\mu}{}_{\nu\alpha}$ and $g_{\mu\alpha,\nu}$ respectively, into pure tensor parts $\Delta^{\mu}{}_{\nu\alpha}$ and $g_{\mu\alpha}|_{\nu}$ and non-covariant,

non-tensor coordinate frame dependent terms $Z^{\lambda}{}_{\mu\nu}$ and $z_{\mu\alpha\nu}$ as

$$\Gamma^{\mu}{}_{\nu\alpha} = (\Delta^{\mu}{}_{\nu\alpha} + Z^{\mu}{}_{\nu\alpha}) \qquad (A.16\text{-a})$$

where

$$Z^{\lambda}{}_{\mu\nu} = 1/2\ \gamma^{\lambda\alpha}(\gamma_{\mu\alpha,\nu} + \gamma_{\nu\alpha,\mu} - \gamma_{\mu\nu,\alpha}) \qquad (A.16\text{-b})$$

and

$$g_{\mu\alpha,\nu} = (g_{\mu\alpha}|_{\nu} + z_{\mu\alpha\nu}) \qquad (A.17\text{-b})$$

where

$$z_{\mu\alpha\nu} = (Z^{\alpha}{}_{\mu\lambda} g_{\alpha\nu} + Z^{\alpha}{}_{\nu\lambda} g_{\alpha\mu}) \qquad (A.17\text{-b})$$

and $\gamma^{\mu\nu}$ is the metric of the flat background spacetime upon which the curvature associated with the real gravitational effects has been imposed. Since the covariant derivative $A_{\mu\nu};\lambda$ of a tensor $A_{\mu\nu}$ remains a tensor in both the ordinary spacetime context and in the bi-metric spacetime context and it does not contain a coordinate frame dependent "z" component.

We see this as follows

$$A_{\mu\nu\,;\,\lambda} = (A_{\mu\nu\,,\,\lambda} - \Gamma^{\alpha}{}_{\mu\lambda} A_{\alpha\nu} - \Gamma^{\alpha}{}_{\nu\lambda} A_{\alpha\mu})$$

$$= (A_{\mu\nu\,|\,\lambda} - \Delta^{\alpha}{}_{\mu\lambda} A_{\alpha\nu} - \Delta^{\alpha}{}_{\nu\lambda} A_{\alpha\mu}) \qquad (A.18)$$

where $\Delta^{\mu}{}_{\nu\alpha} = (\Gamma^{\mu}{}_{\nu\alpha} - Z^{\mu}{}_{\nu\alpha})$.

Hence we see that, starting from the field equations written in the context of tensor g-covariant derivatives, the spacetime tensor quantities appearing in covariant field equations in curved spacetime can be covariantly written in the Relativistic Background Spacetime context using the following procedure: **(a)** replace ordinary non-tensor differentiation (denoted by the subscript comma " , ") by tensor flat background γ-differentiation" (denoted by the subscript bar-symbol " | ") and , **(b)** replace the non-tensor g- Christoffel 3-index symbol $\Gamma^{\mu}{}_{\nu\alpha}$ by the Christoffel 3-index tensor $\Delta^{\mu}{}_{\nu\alpha}$ and, **(c)** replace $(-g)^{1/2} dx^4$ by $(-\kappa)^{1/2} dx^4$ (where $(-\kappa)^{1/2} = (-g)^{1/2}/(-\gamma)^{1/2}$) in the invariant volume element.

However in this context the non-tensor quantities $A_{\mu\nu\,,\,\lambda}$ that are generated by taking the ordinary spacetime derivative of tensor quantities $A_{\mu\nu}$ in curved spacetime can be formally be broken into a tensor part and a non-tensor part as

$$A_{\mu\alpha\,,\,\nu} = (A_{\mu\alpha\,|\,\nu} + z_{\mu\alpha\nu}) \qquad (A.19\text{-}a)$$

where the tensor part $A_{\mu\alpha\,|\,\nu}$ associated with the background spacetime derivative is

$$A_{\mu\alpha\,|\,\nu} = A_{\mu\alpha\,,\,\nu} - (Z^{\alpha}{}_{\mu\lambda} A_{\alpha\nu} + Z^{\alpha}{}_{\nu\lambda} A_{\alpha\mu}) \qquad (A.19\text{-}b)$$

and the non-tensor part $z_{\mu\alpha\nu}$ is given by

$$z_{\mu\alpha\nu} = (Z^{\alpha}{}_{\mu\lambda} A_{\alpha\nu} + Z^{\alpha}{}_{\nu\lambda} A_{\alpha\mu}) \qquad (A.19\text{-}c)$$

Hence we can eliminate the coordinate frame dependent " z-" problem from equations in curved spacetime which contain ordinary spacetime derivatives by formulating those equations entirely in terms of flat background spacetime derivatives.

We can do this by first formulating the theory in the context of a physical spacetime associated with a flat Minkowski background where $A_{\mu\alpha,\nu} = A_{\mu\alpha|\nu}$ and then transforming to a general physical spacetime. In other words we can eliminate this problem by considering all ordinary spacetime derivatives to be flat background Minkowski spacetime derivatives before transforming to a more general set of spacetime coordinates.

We conclude that use of flat background covariant γ-derivatives" of the physical spacetime metric tensor $g_{\mu\nu}$ operationally represents a mathematical language in which the physical geometrical effects of gravitation due to spacetime curvature spontaneously emerge independent of the non-covariant, non-tensor components which are generated by flat spacetime coordinate transformations.

APPENDIX B: EXTENDING SPECIAL RELATIVITY TO GENERAL FLAT COORDINATE
SPACETIMES AND THE NEW BI-METRIC GENERAL RELATIVITY

The physical spacetime associated with Special Relativity is the flat Cartesian inertial Minkowski spacetime frame where all coordinate and non-inertial effects vanish. The metric $\eta_{\mu\nu}$ associated this spacetime is a constant tensor which has the same value for all observers independent of their physical location in the spacetime. However Special Relativity also implies that the following two *fundamental relativistic rules of observation* are valid in the flat, Cartesian, Minkowski spacetime:

a) all physical observations involve local measurements by observers, and

b) all observers measure the local speed of light to be v=c,

Using the rules of General Covariance one can extend Special Relativity in the flat, Minkowski spacetime into the realm of flat non-Cartesian, non-inertial spacetime. However in this case the metric becomes a variable tensor function of spacetime whose value depends on the physical location of observers in the spacetime. Hence in order to obtain a consistent extension of Special Relativity into the realm of flat, non-Cartesian, non-inertial spacetime one must formulate the theory so that the two *fundamental relativistic rules of observation* implied by Special Relativity:

a) all physical observations involve local measurements by observers, and

b) all observers measure the local speed of light to be v=c.

*still remain valid in the flat non-Cartesian, non-inertial spacetime .*

This can be accomplished by taking advantage of the fact that global flat spacetime coordinate transformations contain the inherent freedom to arbitrarily translate the spacetime origin of the observer to spacetime locations which are not necessarily the same as that of the origin of symmetry of the original coordinate transformation. This means that flat spacetime coordinate transformations must be represented in a functional form where the location of the observer is operationally given as part of the physical definition of the transformation . Hence we must define the class of general flat spacetime transformations from a Minkowski spacetime frame to a general flat spacetime frame in an observer-dependent form given by

$$x'^{\mu} = x'^{\mu}(x - x_{obs}) \qquad (B.1)$$

where: a) the spacetime origin of symmetry of the flat spacetime coordinate transformation is located at the origin of the unprimed frame at $x = 0$, b) the spacetime location of the observer in the primed frame is located at the origin $x'_{obs}{}^{\mu}(0) = 0$, while in the unprimed frame it is at $x = x_{obs} \neq 0$.

Now assuming that the unprimed spacetime frame is the Minkowski spacetime the flat $\eta_{\mu\nu}$ metric associated with the spacetime of Special Relativity will transform under the general flat space transformation defined in (1) into an observer-dependent flat background metric tensor $\gamma_{\mu\nu}(x'-x'_{obs})$ where

$$\gamma_{\mu\nu}(x'-x'_{obs}) = (\partial x^\rho / \partial x'^\mu)(\partial x^\sigma / \partial x'^\nu)\eta_{\rho\sigma} \qquad (B.2)$$

and the invariant spacetime interval $ds^2 = \eta_{\mu\nu}dx^\mu dx^\nu$ transforms into the observer-dependent form given by

$$ds'^2(x'-x'_{obs}) = \gamma_{\mu\nu}(x'-x'_{obs})\,dx'^\mu dx'^\nu \qquad (B.3)$$

where since $(-\gamma)^{1/2} = (-\eta)^{1/2} = 1$.

It follows that the extension of Special Relativity into the realm of observer-dependent flat spacetime coordinate transformations defined in (1) satisfies the *fundamental rules of relativistic observation* that : a)) all physical observations involve local measurements by observers, and b) all observers measure the local speed of light to be v=c.

To see this we note that the local coordinate speed of light associated with the observer-dependent the line element (B.3) at the location of the observer where $x'-x'_{obs} = 0$ is given by

$$ds'^2(0) = \gamma_{\mu\nu}(0)\,dx'^\mu dx'^\nu = 0 \qquad (B.4)$$

Now even if the primed frame may be a non-inertial frame where $\partial'_\mu[\gamma(0)_{\alpha\beta}] \neq 0$, the

metric $\gamma_{\mu\nu}(0)$ has the following properties in the local vicinity of the observer:

    a)  $\gamma'_{00}(0) = 1$ and,

    b)  all off-diagonal elements vanish $K \neq J = 1,2,3$, as

$$\gamma'_{0K}(0) = \gamma'_{K0}(0) = 0$$

$$\gamma'_{KJ}(0) = \gamma'_{JK}(0) = 0 \qquad (B.5)$$

Hence (4) becomes

$$ds'^2(0) = dx'^0 dx'^0 + \Sigma_K \gamma'_{KK}(0)\,dx'^K dx'^K) = 0 \qquad (B.6)$$

Dividing (B.6) by $dx'^0 = c\, dt'$ and defining $V'^K = dx'^K / dt'$ we find that the observer at $x' = x'_{obs}$ will locally measure the speed of light to be

$$V' = [\Sigma_K (-\gamma'_{KK})(0)\, v'^K v'^K)]^{1/2} = c \qquad (B.7)$$

*even though the observer is located in a general spacetime frame of reference.*

Hence it follows that the bi-metric form of Einstein General Relativity with its LCC exponential metric expressed as a function of a gravitational potential tensor, contains a new physical paradigm which extends the observer-dependent nature of the Relativity of flat spacetime coordinates to include the requirement that the metric of the curved spacetime should be defined in a observer-dependent manner so that the operational procedure of local spacetime measurements, as seen by an observer in a general spacetime frame which contains gravitational fields due to spacetime curvature, becomes locally defined in a similar manner as that of Special Relativity. In essence the new paradigm requires that an observer at an arbitrary spacetime point $x_{obs}$, in a general frame of reference which contains both the effects of real gravitation due to spacetime curvature and artificial gravitation due to non-inertial coordinate transformations, will locally (i.e. at $x = x_{obs}$) measure the vacuum velocity of light to be $v = c$.

On the basis of the Strong Principle of Equivalence this occurs because: **a)** In addition to the symmetric spacetime metric field $g_{\mu\nu} = g_{\nu\mu}$, there exists a *fundamental* symmetric gravitational potential tensor field $\Phi_{\mu\nu} = \Phi_{\nu\mu}$, which *underlies* the metric effects of real gravitation due to spacetime curvature, and **b)** In the presence of real gravitational fields an observer-dependent exponential metric relationship exists between the $g_{\mu\nu}$ and the mixed tensor $\Phi_\mu{}^\nu$, which allows the operational procedure of local spacetime measurements to be defined for an observer located at spacetime coordinates $x = x_{obs}$ in a similar manner as that of Special Relativity, and for which space and time dilation effects are put on an equal group theoretical footing, given by

$$g_{\mu\nu}(x, x_{obs}) = \exp[(\Phi(x) - \Phi(x_{obs}))_\mu{}^\alpha]\, \gamma_{\alpha\beta}(x - x_{obs})\, \exp[(\Phi^T(x) - \Phi^T(x_{obs}))^\beta{}_\nu] \qquad (B.8)$$

since by definition $\Phi(x_{obs})_\mu{}^\nu = 0$ at the coordinates of the observer which are not necessarily the same as the origin of the spacetime (i.e. $x_{obs} \neq 0$).. Note in the above that:: 1) $g_{\mu\nu}(x, x_{obs})$ depends only on differences in the real gravitational potential tensor field $\Phi(x)_\mu{}^\nu$ and $\Phi(x_{obs})_\mu{}^\nu = 0$, where $\Phi(x)_\mu{}^\alpha \equiv [\phi(x)_\alpha{}^\alpha \delta_\mu{}^\nu - 2\, \phi(x)_\mu{}^\nu]$,

2) $\gamma_{\alpha\beta}(x-x_{obs})$ is the observer dependent metric of the flat background inertial spacetime frame of Special Relativity, and 3) the asymptotic boundary conditions on $\Phi(x)_\mu^\alpha$ are determined by the cosmological boundary conditions on $g_{\mu\nu}(x, x_{obs})$ at spatial infinity which are seen by the observer. On the basis of the above we can distinguish the basic difference between the new bi-metric theory of gravitation and the Einstein theory of Gravitation namely that:

**1)** In the new theory of gravitation the existence of an exponential connection between the metric tensor $g_{\mu\nu}$ and a gravitational potential tensor $\phi_\mu^\nu$ allows the Strong Principle of Equivalence to propagate the observer-dependent property inherent in the flat spacetime metric $\gamma_{\alpha\beta}(x', x'_{obs})$ into the structure of the curved spacetime metric $g'_{\mu\nu}(x', x'_{obs})$.

**2)** In the Einstein theory of gravitation the connection between the metric tensor $g'_{\mu\nu}(x')$ and a gravitational potential tensor does not exist, hence the Strong Principle of Equivalence cannot propagate the observer-dependent property inherent in the flat spacetime metric $\gamma_{\alpha\beta}(x', x'_{obs})$ into the structure of the curved spacetime metric $g'_{\mu\nu}(x')$. .

# APPENDIX C: LOCAL OBSERVER IN A GENERAL BI-METRIC SPACETIME FRAME OF REFERENCE

In a general physical spacetime frame the associated background spacetime metric $\gamma_{\mu\nu}$ is different from the Minkowski background spacetime metric $\eta_{\mu\nu}$ hence $\Sigma^{\lambda}{}_{\mu\nu}$ will in general be nonzero. In this case the contravariant equations of motion for particle of mass $m$ being acted on by gravitational and a non-gravitational force $K^{\mu}$ in this general spacetime frame are

$$K^{\mu}/m = du^{\mu}/ds + \Gamma^{\mu}{}_{\alpha\beta} u^{\alpha} u^{\beta} = du^{\mu}/ds + [\Sigma^{\mu}{}_{\alpha\beta} + \Delta^{\mu}{}_{\alpha\beta}] u^{\alpha} u^{\beta} \tag{C.1}$$

where

$$\Delta^{\lambda}{}_{\mu\nu} = 1/2 \, g^{\lambda\alpha} ( g_{\mu\alpha} |\nu + g_{\nu\alpha} |\mu - g_{\mu\nu} |\alpha ) \tag{C.2-a}$$

$$\Sigma^{\lambda}{}_{\mu\nu} = 1/2 \, \gamma^{\lambda\alpha} ( \gamma_{\mu\alpha},\nu + \gamma_{\nu\alpha},\mu - \gamma_{\mu\nu},\alpha ) \tag{C.2-b}$$

The associated covariant geodesic equations of motion have a more compact form given by

$$K_{\mu}/m = du_{\mu}/ds = 1/2 \, (\gamma_{\alpha\beta},\mu + g_{\alpha\beta} |\mu) u^{\alpha} u^{\beta} \tag{C.3}$$

In the general spacetime frame the observer-dependent exponential metric is given in the bi-metric spacetime context as

$$g_{\mu\nu}(x, x_0) = \exp[(\Phi(x) - \Phi(x_0))_{\mu}{}^{\alpha}] \, \gamma_{\alpha\beta}(x - x_0) \, \exp[(\Phi(x) - \Phi(x_0))^{\beta}{}_{\nu}] \tag{C.4-a}$$

where

$$\Phi(x)_{\mu}{}^{\nu} = (\phi(x)_{\alpha}{}^{\alpha} \delta_{\mu}{}^{\nu} - 2 \phi(x)_{\mu}{}^{\nu}) \tag{C.4-b}$$

and the observer-dependent line element is given by

$$ds^2(x, x0) = g(x, x0)_{\mu\nu} \, dx^\mu \, dx^\nu \tag{C.4-c}$$

Note that in equations (4) above:

**a)** $\Phi(x)_\mu{}^\nu$ is the gravitational potential tensor function, associated with real gravitation due to spacetime curvature, generated by the action of the energy momentum tensor $\tau_\mu{}^\nu$,

**b)** the observer-dependent flat space metric $\gamma(x, x0)_{\mu\nu}$, which is a general solution to the flat space equation $R_{\mu\nu\alpha\beta} = 0$, *appears explicitly* in the exponential metric and differential line element.

Now using (C.4) we can calculate $g_{\alpha\beta \mid \mu}$ and $\Gamma^\mu{}_{\alpha\beta} = \Delta^\mu{}_{\alpha\beta}$ in terms of the quantity $\Phi(x)_\alpha{}^\mu$. For example $g(x, x0)_{\alpha\beta \mid \mu}$ is given (see the end of Appendix D) by

$$g(x, x0)_{\alpha\beta \mid \mu} = 2 \, \Phi_\alpha{}^\rho(x)_{\mid \mu} \, g(x, x0)_{\rho\beta} \tag{C.5}$$

Then the local observer's matter equation of motion is given by the (1) where fictitious and real gravitational forces act in addition to non-gravitational forces as

$$K^\mu / m = du^\mu / ds(x, x0) + \Gamma(x, x0)^\mu{}_{\alpha\beta} \, u^\alpha \, u^\beta$$

$$= du^\mu / ds(x, x0) + [\Sigma(x, x0)^\mu{}_{\alpha\beta} + \Delta(x, x0)^\mu{}_{\alpha\beta}] u^\alpha \, u^\beta \tag{C.1}'$$

while the local observer's line element is given by

$$ds^2(x, x0) = g(x, x0)_{\mu\nu} \, dx^\mu \, dx^\nu \tag{C.4-c}'$$

Then it follows that for an observer in the general spacetime located at the spacetime point x0, the line element for a light ray measured *nonlocally* at the spacetime point x is given by

$$ds^2(x, x0) = g(x, x0)_{\mu\nu} \, dx^\mu \, dx^\nu = 0 \qquad \text{(C.7-a)}$$

where the metric $g_{\mu\nu}$ will now have off-diagonal elements due to spacetime curvature and non-inertial coordinate dependent effects in the general spacetime frame.

However for an observer in the general spacetime located at the spacetime point x0, the line element for a light ray measured *locally* at the spacetime point x0 is

$$\begin{aligned} ds^2(x0,,x0) &= g(x0, x0)_{\mu\nu} \, dx^\mu \, dx^\nu \\ &= \gamma(x0, x0)_{\mu\nu} \, dx^\mu \, dx^\nu = 0 \end{aligned} \qquad \text{(C.7-b)}$$

This implies that in general for a local observer who is in a non-inertial frame at x = x0, but is not freely falling since $[\partial_\mu \gamma(x0, x0)_{\alpha\beta}] \neq 0$ we still have that the non-inertial metric $\gamma(x0, x0)_{\mu\nu}$ has the following properties:

$$\gamma(x0, x0)_{00} = 1,$$
all off-diagonal elements vanishing as:
$$\gamma(x0, x0)_{0K} = \gamma(x0, x0)_{K0} = 0, \quad K = 1, 2, 3$$
$$\gamma(x0, x0)_{KJ} = \gamma(x0, x0)_{JK} = 0, \quad K \neq J = 1, 2, 3 \qquad \text{(C.7-c)}$$

Hence (7-b) implies that a non-freely falling observer at $x' = x'_{obs}$ will locally measure the speed of light to be given by

$$v = [ - \gamma(x0, x0)]_{KK} \, v^K v^K ]^{1/2} = c \qquad (C.8\text{-a})$$

even though as shown in equation (1) 'the local observer is in a non-inertial frame of reference and is not freely falling since

$$K^\mu / m = du^\mu / ds(x0, x0) + \Gamma(x0, x0)^\mu{}_{\alpha\beta} \, u^\alpha u^\beta$$

$$= du^\mu / ds(x0, x0) + [\Sigma(x0, x0)^\mu{}_{\alpha\beta} + \Delta(x0, x0)^\mu{}_{\alpha\beta}] u^\alpha u^\beta \qquad (C.8\text{-b})$$

here $\Sigma(x0, x0)^\mu{}_{\alpha\beta}$ generates the fictitious gravitational forces and $\Delta(x0, x0)^\mu$ generates the real gravitational forces. Note in addition that we also have, but do not necessarily have to use, the option to locally transform the observer-dependent flat space metric $\gamma(x0, x0)_{\mu\nu}$ into a locally Minkowskian form such that $\gamma(x0, x0)_{\rho\sigma} = \eta_{\mu\nu}$ even though the first derivatives $\gamma(x, x0)_{\mu\nu, \lambda}$ will not necessarily vanish at $x=x0$ this general flat spacetime frame.

# APPENDIX D: LOCAL OBSERVER IN A FREE-FALL BI-METRIC SPACETIME FRAME OF REFERENCE

Now to define a "local free-fall frame of reference" at the spacetime point $x = x_0$ let us consider a local non-linear non-inertial co-ordinate transformation given to lowest order about the general observer spacetime point $x = x0$ defined by

$$x'^{\mu} = x^{\mu} - 1/2\, [(A^{\mu}{}_{\alpha\beta})(x^{\alpha} - x_0{}^{\alpha})(x^{\beta} - x_0{}^{\beta})] \qquad \text{(D.1-a)}$$

$$dx'^{\mu} = dx^{\mu} - [(A^{\mu}{}_{\alpha\beta})(x^{\alpha} - x_0{}^{\alpha})]\, dx^{\beta} \qquad \text{(D.1-b)}$$

and its inverse transformation given by

$$x^{\mu} = x'^{\mu} + 1/2\, [(A^{\mu}{}_{\alpha\beta})(x^{\alpha} - x_0{}^{\alpha})'\,(x^{\beta} - x_0{}^{\beta})'] \qquad \text{(D.1-c)}$$

$$dx^{\mu} = dx'^{\mu} + [(A^{\mu}{}_{\alpha\beta})(x^{\alpha} - x_0{}^{\alpha})']\, dx'^{\beta} \qquad \text{(D.1-d)}$$

Then at the spacetime point $x' = x0$ of the free fall observer both the gravitational metric $g_{\mu\nu}$ and the flat background spacetime metric $\gamma_{\mu\nu}$ will both locally transform as tensors under this local non-inertial transformation to become

$$g'_{\mu\nu} = (dx^{\rho}/dx^{\mu}{}')(dx^{\sigma}/dx^{\nu}{}')\, g_{\rho\sigma} \qquad \gamma'_{\mu\nu} = (dx^{\rho}/dx^{\mu}{}')(dx^{\sigma}/dx^{\nu}{}')\, \gamma_{\rho\sigma} \qquad \text{(D.2-a)}$$

$$ds'^{2} = g'_{\mu\nu}\, dx'^{\mu}\, dx'^{\nu} \qquad d\sigma'^{2} = \gamma'_{\mu\nu}\, dx'^{\mu}\, dx'^{\nu} \qquad \text{(D.2-b)}$$

$$\Sigma'^{\lambda}{}_{\mu\nu} = [\{(dx^{\sigma}/dx^{\mu}{}')(dx^{\tau}/dx^{\mu}{}')(dx^{\lambda}{}'/dx^{\rho})\, \Sigma^{\rho}{}_{\sigma\tau}\}$$
$$+ (d^{2}x^{\sigma}/dx^{\mu}{}'\, dx^{\nu}{}')(dx^{\lambda}{}'/dx^{\sigma})]$$
$$= [\{(dx^{\sigma}/dx^{\mu}{}')(dx^{\tau}/dx^{\mu}{}')(dx^{\lambda}{}'/dx^{\rho})\, \Sigma^{\rho}{}_{\sigma\tau}\} + A^{\mu}{}_{\alpha\beta} \qquad \text{(D.2-c)}$$

$$\Delta'^{\lambda}{}_{\mu\nu} = (dx^{\sigma}/dx^{\mu}{}')(dx^{\tau}/dx^{\mu}{}')(dx^{\lambda}{}'/dx^{\rho})\, \Delta^{\rho}{}_{\sigma\tau} \qquad \text{(D.2-d)}$$

$$\Gamma'^{\lambda}{}_{\mu\nu} = [\{(dx^{\sigma}/dx^{\mu}{}')(dx^{\tau}/dx^{\mu}{}')(dx^{\lambda}{}'/dx^{\rho})\, (\Sigma^{\rho}{}_{\sigma\tau} + \Delta^{\rho}{}_{\sigma\tau})\}$$
$$+ \{(d^{2}x^{\sigma}/dx^{\mu}{}'\, dx^{\nu}{}')(dx^{\lambda}{}'/dx^{\sigma})\}]$$
$$= [\,(\Sigma'^{\rho}{}_{\sigma\tau} + \Delta'^{\lambda}{}_{\mu\nu}) + A^{\lambda}{}_{\mu\nu}] \qquad \text{(D.2-e)}$$

The mass of the particle is a scalar so that $m' = m$, while the non-gravitational force $K^\mu$ transforms like a tensor as

$$K'^\mu = (dx'^\mu / dx_\nu)\, K^\nu \qquad (D.2\text{-}f)$$

However since (1) are tensor equations then at the spacetime point $x0' = x0$ they are

$$K'^\mu / m = du'^\mu / ds(x'0, x'0)' = \Gamma'^\mu{}_{\nu\alpha}(x'0, x'0)\, u'^\alpha u'^\beta ,$$

$$= [(\Sigma'^\mu{}_{\nu\alpha}(x'0, x'0) + \Delta'^\mu{}_{\nu\alpha}(x'0, x'0)) + A^\mu{}_{\nu\alpha}]\, u'^\alpha u'^\beta \qquad (D.3)$$

The local free-fall frame of reference is define by choosing the local non-inertial transformation (9) so that the non-inertial forces, generated by the quantity

$\Sigma'^\mu{}_{\nu\alpha}(x'0, x'0) + A^\mu{}_{\nu\alpha}$ at the spacetime point $x0' = x0$, **compensate and cancel out** the real gravitational forces $\Delta'^\mu{}_{\nu\alpha}(x'0, x'0)$ then the total force vanishes since $\Gamma'^\mu{}_{\nu\alpha}(x0, x0) = 0$. Then in (D.3)

$$A^\mu{}_{\alpha\beta} = -[\Sigma'^\mu{}_{\nu\alpha}(x0, x0) + \Delta'^\mu{}_{\nu\alpha}(x0, x0)] \qquad (D.4\text{-}a)$$

$$A^\mu{}_{\alpha\beta} = -[\Sigma'^\mu{}_{\nu\alpha}(x0, x0) + \Phi'^\mu{}_\alpha{}_{|\beta}(x0) + \Phi'^\mu{}_\beta{}_{|\alpha}(x0)$$

$$- 1/2(\Phi'(x0)_\alpha{}^{\lambda|\mu}\gamma(x0, x0)_{\lambda\beta} + \Phi'(x0)_\beta{}^{\lambda|\mu}\gamma(x0, x0)_{\lambda\alpha}) \qquad (D.4\text{-}b)$$

and for the free fall observer at $x'=x0$

$$g'_{\mu\nu}(x0, x0) = \gamma(x0, x0)_{\mu\nu}$$

$$ds'^2(x0, x0) = \gamma(x0, x0)_{\mu\nu}\, dx'^\mu dx'^\nu$$

$$\Gamma'^\mu{}_{\nu\alpha}(x0, x0) = 0 \qquad (D.4\text{-}c)$$

Where again we also have, but do not necessarily have to use, the option to locally transform the observer-dependent flat space metric $\gamma'(x0, x0)_{\mu\nu}$ into a locally Minkowskian form such that $\gamma'(x0, x0)_{\rho\sigma} = \eta_{\mu\nu}$ even though the first derivatives $\gamma'(x0, x0)_{\mu\nu, \lambda}$ will not necessarily vanish at $x=x0$ in this general flat spacetime frame. Under these conditions the local effects of gravitation vanish in equation (D.3) and it becomes locally identical to that of Special Relativity where

$$K'^{\mu} / m = du'^{\mu} / ds' \,(x0, x0) \qquad (D.4\text{-}d)$$

Hence using (D.3) we see that the free fall Special Relativity correspondence limit, in the bi-metric spacetime can be defined at each spacetime point $x = x_0$ by a local non-inertial coordinate transformation given in terms of the flat background derivatives of the gravitational potential tensor function $\Phi_\mu{}^\nu = (\phi_\alpha{}^\alpha \delta_\mu{}^\nu - 2\phi_\mu{}^\nu)$ evaluated to first order at the spacetime point $x'0 = x0$ given by

$$dx'^{\mu} = dx^{\mu} - \{\Sigma'{}^{\mu}{}_{\nu\alpha}(x0, x0) + \Phi_\alpha{}^\mu{}_{|\beta}(x0) + \Phi_\beta{}^\mu{}_{|\alpha}(x0)$$

$$- 1/2\, (\Phi'(x0)_\alpha{}^{\lambda|\mu} \gamma'(x0, x0)_{\lambda\beta} + \Phi'(x0)_\beta{}^{\lambda|\mu} \gamma'(x0, x0)_{\lambda\alpha}\} (x^\nu - x0^\nu)\, dx^\alpha \qquad (D.4\text{-}e)$$

This means that the **"instantaneous rest Lorentz frame observer"** which Special Relativity uses to define the relativistic particle equations of motion of the form

$$K'^{\mu} = m\,(du'^{\mu} / ds') \qquad (D.5\text{-}a)$$

where

$$ds'^2(x0, x0) = \gamma'(x0, x0)_{\mu\nu} dx'^{\mu} dx'^{\nu} \quad (D.5\text{-b})$$

is actually the **"local freely falling observer"** defined in the context of the bi-metric space-time by the free fall transformation (D.4-e). Hence this procedure defines the free fall Special Relativty correspondence limit of the new bi-metric General Relativity theory.

Hence we see that the observer-dependent metric measured by the **"local freely falling instantaneous rest Lorentz frame observer"** given by

$$ds'^2(x0, x0) = \gamma'(x0, x0)_{\mu\nu} dx'^{\mu} dx'^{\nu} \quad (D.6\text{-a})$$

is the same as that of the **local non-inertial rest Frame observer** given by

$$ds^2(x0, x0) = \gamma(x0, x0)_{\mu\nu} dx^{\mu} dx^{\nu} \quad (D.6\text{-b})$$

As an elementary example of this consider an elevator in free-fall in the gravitational field of the earth. In the $v <<< c$ low velocity limit the only non-zero terms which contribute to the gravitational force term $[\Sigma^{\mu}_{\nu\alpha}{}' + \Delta^{\mu}_{\nu\alpha}{}']$ in the geodesic equation are the terms associated with a constant acceleration in the z-direction given by

$$a^3 = A^3{}_{00} = [a/c^2] = -[\Phi_0{}^3{}_{,0}(x_O) + \Phi_0{}^3{}_{,0}(x_O) - \Phi_{00}{}^{,3}(x_O)] \quad (D.7\text{-a})$$

$$cdt' = cdt, \quad dx' = dx, \quad dy' = dy, \quad dz' = [dz - A^3{}_{00}(ct) cdt] = [dz - at\, dt] \quad (D.7\text{-b}$$

$$\Sigma^3{}_{00}{}' = A^3{}_{00} = a^k/c^2 = -\Delta^3{}_{00}(0) = -\Phi(0)_0{}^0{}_{,3} = -g/c^2 \quad (D..7\text{-c})$$

in the geodesic equations of motion leading to the Special Relativity equation of motion

$$K'^{\mu} = m \, du'^{\mu} / ds' \qquad (D.7\text{-}d)$$

which again shows how the kinematic force generated by the local non-inertial coordinate transformation to free-fall frame locally compensates the actual gravitational force . Now using a general flat space-time transformation let us transform from the Inertial Cartesian Minkowski coordinates associated with the flat background space-time metric $\eta_{\mu\nu}$, and the curved metric $g_{\mu\nu}$, to a general set of space-time coordinates associated with the flat background space-time metric $g'_{\mu\nu}$ and the curved metric $\gamma'_{\mu\nu}$.

Since $\gamma'_{\mu\nu}$ is different from $\eta_{\mu\nu}$ the flat Christoffel symbols $\Sigma'^{\lambda}_{\mu\nu}$ will in general be nonzero. In this case (now dropping the prime notation) the contravariant equations of motion for proper mass density $\sigma$ being acted on by gravitational and a non-gravitational force $K^{\mu}$ in this general space-time frame is are

$$K^{\mu} = \sigma \{ du^{\mu}/ds + \Gamma^{\mu}_{\alpha\beta} u^{\alpha} u^{\beta} \} = \sigma \{ du^{\mu}/ds + [\Sigma^{\mu}_{\alpha\beta} + \Delta^{\mu}_{\alpha\beta}] u^{\alpha} u^{\beta} \} \qquad (D.8\text{-}a)$$

which can be also written in the form

$$\sigma(du^\mu/ds + \Delta^\mu{}_{\alpha\beta} u^\alpha u^\beta) = (K^\mu - \sigma\Sigma^\mu{}_{\alpha\beta} u^\alpha u^\beta) \tag{D.8-b}$$

where

$$\Sigma^\lambda{}_{\mu\nu} = 1/2\ \gamma^{\lambda\alpha}(\gamma_{\mu\alpha,\nu} + \gamma_{\nu\alpha,\mu} - \gamma_{\mu\nu,\alpha}) \tag{D.8-c}$$

$$\Delta^\lambda{}_{\mu\nu} = (\Gamma^\lambda{}_{\mu\nu} - \Sigma^\lambda{}_{\mu\nu})$$

$$= 1/2\ [\ g^{\lambda\alpha}(g_{\mu\alpha,\nu} + g_{\nu\alpha,\mu} - g_{\mu\nu,\alpha}) - \gamma^{\lambda\alpha}(\gamma_{\mu\alpha,\nu} + \gamma_{\nu\alpha,\mu} - \gamma_{\mu\nu,\alpha})\ ]$$

$$= 1/2\ g^{\lambda\alpha}(g_{\mu\alpha}|\nu + g_{\nu\alpha}|\mu - g_{\mu\nu}|\alpha) \tag{D.8-d}$$

and

$$g_{\mu\nu}|\lambda = g_{\mu\nu,\lambda} - \Sigma^\alpha{}_{\mu\lambda} g_{\alpha\nu} - \Sigma^\alpha{}_{\nu\lambda} g_{\alpha\mu} \tag{D.8-e}$$

In the general bi-metric space-time frame the exponential metric is given by

$$g_{\mu\nu} = \exp(\Phi)_\mu{}^\alpha]\ \gamma_{\alpha\beta}\ \exp(\Phi^T)^\beta{}_\nu] \tag{D.9-a}$$

where

$$\Phi_\mu{}^\nu = (\phi_\alpha{}^\alpha\ \delta_\mu{}^\nu - 2\ \phi_\mu{}^\nu) \tag{D.9-b}$$

and the line element is given by

$$ds^2 = g_{\mu\nu}\ dx^\mu\ dx^\nu \tag{D.9-c}$$

Then inside of the matter equation of motion (1-b) in the general space-time frame

$$\sigma(du^\mu/ds + \Delta^\mu{}_{\alpha\beta} u^\alpha u^\beta) = (K^\mu - \sigma\Sigma^\mu{}_{\alpha\beta} u^\alpha u^\beta) \tag{D.9-b}$$

where the bi-metric gravitational field equation is given by

$$-1/2\, G_\mu{}^\nu = (4\pi G / c^4)\, \tau_\mu{}^\nu \qquad \text{(D.10-b)}$$

and the Bianchi-Freud Identity is

$$[\tau_\mu{}^\nu]_{;\nu} \equiv 0 \qquad \text{(D.11-c)}$$

Now equation (D.9-b) will describe a class of local "free-fall" observers if $K^\mu = 0$ and

(D.9-b) describes trajectories such that $\Sigma^\mu{}_{\alpha\beta} = -\Delta^\mu{}_{\alpha\beta} \neq 0$ at every point on these

trajectories. Then the non-inertial gravitational forces associated with $\Sigma^\mu{}_{\alpha\beta} \neq 0$

cancel out the local real gravitational forces associated with $\Delta^\mu{}_{\alpha\beta} \neq 0$ so that the local observable force associated with $\Gamma^\mu{}_{\alpha\beta}$ vanishes locally in these observer's equations of motion as $\Gamma^\mu{}_{\alpha\beta} = 0$ implying that the class of "free-fall" observer's local equation of motion is given by $\sigma du^\mu / ds = 0$.

## APPENDIX E:   INTERACTIVE N-BODY SOLUTIONS IN BI-METRIC EINSTEIN GENERAL RELATIVITY

In the absence of pressure $P = 0$ the contravariant geodesic equations of motion for matter are given by

$$\sigma \, du^\mu / ds \;=\; -\Gamma^\mu{}_{\alpha\beta} \, (\sigma u^\alpha u^\beta) \;=\; -(Z^\mu{}_{\alpha\beta} + \Delta^\mu{}_{\alpha\beta})\,(\sigma u^\alpha u^\beta) \qquad (E\text{-}1)$$

where in the inertial physical space-time associated with the Minkowski background space-time $Z^\mu{}_{\alpha\beta} = 0$ and the Christoffel symbol $\Gamma^\mu{}_{\alpha\beta}$ is equal to the Christoffel tensor $\Delta^\mu{}_{\alpha\beta}$

$$\Gamma^\mu{}_{\alpha\beta} \;=\; 1/2 \, g^{\mu\rho} (g_{\alpha\rho,\beta} + g_{\beta\rho,\alpha} - g_{\alpha\beta,\rho}) \;=\; \Delta^\mu{}_{\alpha\beta} \qquad (E\text{-}2)$$

We recall that in the new bi-metric gravitational the field equations the metric $g_{\mu\nu}$ is defined in terms of the symmetric gravitational potential tensor matrix

$\Phi_\mu{}^\nu \equiv \Phi$  (using space-time 4x4 matrix notation) as

$$g_{\mu\nu} \;=\; \exp[(\Phi)_\mu{}^\alpha] \; \eta_{\alpha\beta} \; \exp[(\Phi)^\beta{}_\nu] \qquad (E\text{-}3)$$

where $\Phi_\mu{}^\nu$ is a function of the gravitational potential tensor $\phi_\mu{}^\nu$ given by

$$\Phi \equiv \Phi_\mu{}^\nu \;=\; (\mathrm{Tr}(\phi)\,\delta_\mu{}^\nu - 2\phi_\mu{}^\nu) \qquad (E\text{-}4)$$

and in the matrix exponential is defined in terms of the power series

$$[\exp(\Phi)]_\mu{}^\nu \equiv \delta_\mu{}^\nu + \Sigma_{N=1,2,\ldots} \, [(\Phi)^N]_\mu{}^\nu / N! \qquad (E\text{-}5)$$

Now from equation (28) we have that

$$\frac{1}{2} g^{\nu\rho} g_{\mu\rho,\alpha} = \Phi_\mu{}^\nu{}_{,\alpha} \qquad (E\text{-}6\text{-}a)$$

and

$$\frac{1}{2} g^{\mu\rho} g_{\alpha\beta,\rho} = \frac{1}{2} g^{\mu\rho} (g_{\sigma\alpha} \Phi_\beta{}^\sigma{}_{,\rho} + g_{\sigma\beta} \Phi_\alpha{}^\sigma{}_{,\rho}) \qquad (E\text{-}6\text{-}b)$$

Hence we have that the Christoffel symbol $\Gamma^\mu{}_{\alpha\beta}$ can be written as

$$\Gamma^\mu{}_{\alpha\beta} = \Phi_\alpha{}^\mu{}_{,\beta} + \Phi_\beta{}^\mu{}_{,\alpha} - \frac{1}{2}(g_{\sigma\alpha} \Phi_\beta{}^{\sigma,\mu} + g_{\sigma\beta} \Phi_\alpha{}^{\sigma,\mu}) \qquad (E\text{-}6\text{-}c)$$

Hence the contravariant geodesic equation of motion can be written in terms of the derivatives of $\Phi_\alpha{}^\mu$ as

$$\sigma\, du^\mu/ds = -\Gamma^\mu{}_{\alpha\beta}(\sigma u^\alpha u^\beta) = -[\Phi_\alpha{}^\mu{}_{,\beta} + \Phi_\beta{}^\mu{}_{,\alpha} - \frac{1}{2}(g_{\sigma\alpha} \Phi_\beta{}^{\sigma,\mu} + g_{\sigma\beta} \Phi_\alpha{}^{\sigma,\mu})](\sigma u^\alpha u^\beta) \qquad (E\text{-}7)$$

Now if the effects of gravitational radiation are very small, the gravitational N-body problem can be formulated in bi-metric General Relativity by assuming and that energy momentum tensor can be broken up into N distinct parts given by $\tau_\mu{}^\nu = \Sigma_{1,2,\ldots N}\, \tau(N)_\mu{}^\nu$. Then the gravitational potential tensor $\phi_\mu{}^\nu$ will break up into N distinct parts given by $\phi_\mu{}^\nu = \Sigma_{1,2,\ldots N}\, \phi(N)_\mu{}^\nu$ under the condition that within the covariant volume integral of the bi-metric gravitational energy-momentum tensor $<U_\mu{}^\nu>$., (which is bi-linear in the products of bi-metric covariant derivatives of the gravitational potential tensor $\phi_\mu{}^\nu{}_{|\alpha}$) the covariant volume integral of the bi-linear cross-terms in $<U(NM)_\mu{}^\nu>$ are negligible compared to the covariant volume integral of the diagonal terms $<U(NN)_\mu{}^\nu>$ .so that the covariant volume integral of the bi-metric gravitational energy-momentum tensor breaks up into N distinct parts as $<U_\mu{}^\nu> = \Sigma_{1,2,\ldots N} <U(NN)_\mu{}^\nu>$.

Then choosing the Harmonic tensor gauge where $\phi_\mu{}^\nu{}_{|\nu} = 0$ <-----> $g^{\mu\nu}{}_{|\nu}$ we find in this approximation that the gravitational potential tensor $\Phi_\mu{}^\nu$ can be represented by a sum N distinct potentials given by the gravitation potential tensor $\phi_\mu{}^\nu = \Sigma_{1,2,...N}\ \phi_\mu{}^\nu(N)$ which when inserted into the Einstein equation

$$\{\exp[Tr(2\phi)]\ (\phi_\mu{}^{\nu|\alpha} - \phi_\mu{}^{\alpha|\nu})\}_{|\alpha} = \{\exp[Tr(2\phi)]\ (\tau_\mu{}^\nu + U_\mu{}^\nu)\} \qquad \text{(E-8-a)}$$

implies that

$$\Sigma_{1,2,...N}\ \{\exp[Tr(2\phi)]\ (\phi(N)_\mu{}^{\nu|\alpha} - \phi(N)_\mu{}^{\alpha|\nu})\}_{|\alpha}$$
$$- \{\exp[Tr(2\phi)]\ (\tau(N)_\mu{}^\nu + U(NN)_\mu{}^\nu)\} = 0 \qquad \text{(E-8-b)}$$

where

$$U(NN)_\mu{}^\nu = [1/4\ \delta_\mu{}^\nu[\ Tr\ (\Phi(N)^{|\alpha})\ Tr\ (\Phi(N)_{|\alpha})] - Tr\ (\Phi(N)_{|\mu})\ (\Phi(N)^{T|\nu})] \qquad \text{(E-8-c)}$$

*Now since this tensor equation valid for arbitrary values of spacetime coordinates x, this implies that each $\phi(N)_\mu{}^\nu$ obeys its own Einstein equation*

$$\{\exp[Tr(2\phi)]\ (\phi(N)_\mu{}^{\nu|\alpha} - \phi(N)_\mu{}^{\alpha|\nu})\}_{|\alpha} - \{\exp[Tr(2\phi)]\ (\tau(N)_\mu{}^\nu + U(NN)_\mu{}^\nu)\} \qquad \text{(E-8-d)}$$

and since in this approximation it $\Phi_\mu{}^\nu = \Sigma_{1,2,...N}\ \Phi_\mu{}^\nu(N$ then it follows that the N-body metric tensor in the bi-metric form of Einstein General Relativity is given by

$$g^{\alpha\beta} = \exp(\Sigma_{1,2,...M}\ \Phi(M)^\alpha{}_\rho)\ \eta^{\rho\sigma}\ \exp(\Sigma_{1,2,...M}\ \Phi^T(M)_\sigma{}^\beta) \qquad \text{(E-8-e)}$$